\documentclass[pre,10pt,amsmath,amssymb]{revtex4}
\usepackage{graphicx}

\usepackage{amsmath,amsfonts,amssymb,bm}

\begin{document}

\title{Anomalous diffusion in systems driven by the stable L\'evy noise 
with a finite noise relaxation time and inertia}

\author
{Tomasz Srokowski}

\affiliation{
 Institute of Nuclear Physics, Polish Academy of Sciences, PL -- 31-342
Krak\'ow,
Poland }

\date{\today}

\begin{abstract}
Dynamical systems driven by a general L\'evy stable noise are considered. 
The inertia is included and the noise, represented by a generalised 
Ornstein-Uhlenbeck process, has a finite relaxation time. A general 
linear problem (the additive noise) is solved: the resulting distribution 
converges with time to the distribution for the white-noise, massless case. 
Moreover, a multiplicative noise is discussed. It can make 
the distribution steeper and the variance, which is finite, depends sublinearly 
on time (subdiffusion). For a small mass, a white-noise limit corresponds 
to the Stratonovich interpretation. On the other hand, the distribution tails 
agree with the It\^o interpretation if the inertia is very large. An escape time 
from the potential well is calculated. 

\end{abstract} 


\maketitle


\section{Introduction}

A prominent feature of the stable L\'evy processes is the existence of 
algebraic, long tails of the probability distributions of the  
form $|x|^{-\alpha-1}$, where $0<\alpha<2$ is a stability index. 
As a consequence, the moments, in particular the variance, are divergent. 
The diffusion process 
in such systems is called an "accelerated diffusion" and the relative 
transport rate may be described by a time-dependence of the fractional 
moments instead of the variance. However, there are indications 
that in some physical problems the distribution tails fall faster than 
for the pure L\'evy flight. In the field of the economic research, the 
indexes 2.5 -- 4 were observed in financial data \cite{sta}; it has been 
suggested that such values of the index arise when the trading behaviour is 
performed in an optimal way \cite{gab}.  The probability distributions of 
the hydraulic conductivity in the porous media seem to obey the power law 
with the index 3.5, while the atmospheric turbulence studies 
yield even larger index for the wind field  \cite{sche}. Such slowly falling 
algebraic tails are predicted by the Langevin equation with the L\'evy stable 
noise -- in a sense of the stationary solution -- when one introduces 
an appropriate deterministic potential. It has been 
demonstrated by Chechkin et al. \cite{che} that the stationary distribution 
tails of the form $x^{-\alpha-2m-1}$ result from the potential $\sim x^{2m+2}$. 
Also temporal characteristics of the system may influence the asymptotic shape 
of the distribution. This happens if, for a jumping process, 
long jumps are penalised by a short waiting time. The finite variance is observed 
for such jumping processes as the L\'evy walk \cite{kla} and the kangaroo process 
with a L\'evy distributed jumping size \cite{srkan}. 

The L\'evy stable processes are often connected with complex phenomena for which 
the power-law shape of the distribution tails \cite{newm} is typical, 
as well as a complicated structure of the medium. It is the case for the porous 
media, plasmas and fractal (multifractal) structures \cite{sche,park}. 
Therefore, a nonhomogeneity must often be taken 
into account in a dynamical description, both as a deterministic potential 
and as a multiplicative noise. Descriptions of the diffusion on fractals 
involve the variable, power-law diffusion coefficient \cite{osh,met4}. 
Also the other topologically complicated systems with long jumps, the folded polymers, 
require a variable diffusion coefficient to describe the transport \cite{bro}. 
Moreover, formalisms with the multiplicative L\'evy noise can describe 
the second order phase transitions \cite{manor} and 
the dynamics of two competing species \cite{cogn}. 
A nonlinearity of the Langevin equation makes the stochastic process different from 
the pure L\'evy motion. In particular, variance may be finite for a system 
driven by the multiplicative L\'evy noise \cite{sro1}. Generally, 
the variance rises not only linearly with time but also faster or slower than that, 
i.e. the diffusion may be anomalous. In the case of Ref.\cite{sro1}, 
motion is subdiffusive. The above approach includes the white noise. However, 
a Markovian description of a realistic system is an idealisation, valid only if 
the time scale of fast variables is short, compared to the time scale of 
the process variable. A procedure of the fast variables elimination produces 
correlations: they are present even if the original system is Markovian \cite{hae}. 
It has been demonstrated for the Gaussian noise that 
characteristic time scales of the fast variables are important even if the variables 
themselves are eliminated \cite{ter}; this finding suggests using 
a coloured noise in a stochastic description rather than the white noise. 
Effects related to the correlations are important, for example, for such problems as fluctuations 
of a dye laser light \cite{sho} and a narrowing of the magnetic resonance lines \cite{kub}. 
Importance of the finite correlation time for noise-induced phase transitions was 
emphasised in Ref.\cite{mang}; an increase of that time favours disorder and
prevents the formation of an ordered state. 
Introducing the white noise as a limit of the finite correlation time means that 
the stochastic integral should be interpreted in a Stratonovich sense \cite{won}. 
On the other hand, effects of the finite inertia should be taken into account. 
If the relaxation time associated with the inertia is large compared to 
the correlation time, the It\^o interpretation comes into play \cite{kup}. 
That effect of the inertia, opposite to the correlations, was demonstrated in Ref.\cite{san}: 
it modifies the front propagation by suppressing the external multiplicative, 
white noise influence on the velocity of fronts. 

The It\^o-Stratonovich dilemma becomes especially interesting for $\alpha<2$ since then -- 
when we consider the white L\'evy noise and neglect the inertia -- the very 
existence of the variance depends on the particular interpretation of the stochastic 
integral. This problem is important for the diffusion since the infinite variance, 
which means the infinite propagation speed, is unphysical in most cases. 
How do the finite noise relaxation time and the inertia modify slope of 
the distribution? We address this question in the present paper and discuss consequences 
for the diffusion. In Sec.II a linear problem involving the additive noise is 
considered. Sec.III is devoted to the multiplicative noise; both limiting, analytically 
solvable cases and numerical solutions are discussed. Moreover, the escape time from 
a potential well is calculated. Results are summarised in Sec.IV.

\section{Additive noise}

We consider a linear problem which is defined by the following 
system of the Langevin equations for $x$, $v$ and $\xi$: 
\begin{eqnarray}
\label{la}
m\dot v(t)&=&f_0-\beta v(t)-\lambda x+\gamma\xi(t)\nonumber\\
\dot x(t)&=&v(t)\nonumber\\
d\xi(t)&=&-\gamma\xi(t)dt+dL(t), 
\end{eqnarray}
where $\beta$ is a damping coefficient. The stochastic force, 
$dL(t)$, is the symmetric and stable L\'evy process, characterised 
by the stability index $\alpha\in(0,2]$. Special cases of the system (\ref{la}) 
were considered by several authors. The velocity distribution for the white 
noise without a potential was obtained in Ref.\cite{west}, the linear 
force case was discussed in Ref.\cite{jes} and the white noise case with 
the inertia for $\alpha=1$ in Ref.\cite{garb}. Moreover, the asymmetric L\'evy 
distribution was introduced in Ref.\cite{yan}. On the other hand, stochastic 
collision models may lead to the L\'evy statistics. The Fokker-Planck equation 
with the additive noise predicts, in the limit of small mass, 
an equilibrium in the form of the L\'evy distribution \cite{barkai}. 
The case $\alpha=2$ corresponds 
to the normal distribution. Then the third equation (\ref{la}) describes 
the standard Ornstein-Uhlenbeck process with the covariance 
\begin{equation}
\label{ouc}
\langle \xi(t)\xi(0)\rangle=\langle \xi^2(0)\rangle {\hbox{e}}^{-\gamma t};
\end{equation}
therefore $\gamma$ determines a correlation time, $1/\gamma$. A generalisation 
of the Ornstein-Uhlenbeck process for $\alpha<2$ implies an infinite 
covariance for any time. However, the parameter $\gamma$ can still estimate 
the noise relaxation time. One can modify the covariance definition \cite{emb,elia} 
to get a convergent quantity which behaves with time similar to Eq.(\ref{ouc}). 
On the other hand, the covariance becomes finite when one introduces a truncation 
of the L\'evy distribution \cite{act}. Properties 
of such a dynamical system are similar to the system without the truncation 
for an arbitrarily large time \cite{man} and the parameter $\gamma$ measures 
the correlation time. Values of the process $dL(t)$ are given by 
the characteristic function ${\widetilde p}(k)=\exp(-K^\alpha|k|^\alpha)$ 
$(K>0)$. The fractional Fokker-Planck equation 
\begin{equation}
\label{frace}
\frac{\partial}{\partial t}p=-v\frac{\partial}{\partial x}p
-\frac{1}{m}(f_0-\beta v-\lambda x+\gamma \xi)\frac{\partial}{\partial v}p
+\gamma\frac{\partial}{\partial \xi}(\xi p)+
\frac{\beta}{m}p+K^\alpha\frac{\partial^\alpha}{\partial |\xi|^\alpha}p
\end{equation}
determines the probability density distribution $p(x,v,\xi;t|x_0,v_0,\xi_0;0)$ 
and the fractional Weyl derivative is defined by its Fourier transform, 
${\cal F}[\frac{\partial^\alpha}{\partial |x|^\alpha}f(x)]=-|k|^\alpha{\widetilde f}(k)$. 
We will evaluate the density of $x$, $p(x,t)$, directly from the stochastic 
equation (\ref{la}). We restrict our analysis to the case of a relatively weak 
potential; more precisely: let $\beta^2/m^2-4\lambda/m\equiv \Delta^2>0$. 
First, we need to evaluate the stochastic trajectory $x(t)$. The solution of 
Eq.(\ref{la}) produces the result 
\begin{equation}
\label{xodt}
x(t)=f_1+\int_0^tf_2(t')L(t')dt', 
\end{equation} 
where 
\begin{equation}
\label{f1}
f_1=\frac{2v_0}{\Delta}{\hbox{e}}^{-\beta t/2m}\sinh\frac{\Delta}{2}t+
\frac{2f_0}{m\Delta}\left[\frac{m\Delta}{2\lambda}+\exp(-\frac{1}{2}\frac{\beta}{m}t)\left(
\frac{\exp(-\frac{\Delta}{2}t)}{\beta/m+\Delta}-\frac{\exp(\frac{\Delta}{2}t)}{\beta/m-\Delta}\right)\right]
\end{equation}
and 
\begin{equation}
\label{f2}
f_2(t')=\frac{\gamma}{m\Delta}{\hbox{e}}^{-\gamma(t-t')}\left[
\frac{\exp[(\gamma-\frac{\beta}{2m}+\frac{\Delta}{2})(t-t')]}{\gamma-\beta/2m+\Delta/2}-
\frac{\exp[(\gamma-\frac{\beta}{2m}-\frac{\Delta}{2})(t-t')]}{\gamma-\beta/2m-\Delta/2}+
\frac{m\Delta}{m\gamma^2-\gamma\beta+\lambda}\right]. 
\end{equation}
Two simple special cases are distinguished. In the absence of inertia ($m=0$), 
we have an adiabatic problem of the particle subjected to the linear force and 
the coloured L\'evy noise. Then Eq.(\ref{la}) yields 
\begin{equation}
\label{bezm}
f_2(t')=\frac{\gamma}{\lambda-\beta\gamma}\left({\hbox{e}}^{\lambda t/\beta-\gamma(t-t')}-
{\hbox{e}}^{\lambda t'/\beta}\right). 
\end{equation}
Secondly, for the case of a free-particle ($\lambda=0$) we obtain 
\begin{equation}
\label{bezl}
f_2(t')=\frac{1}{\beta}+\left(\frac{1}{\gamma-\beta/m}-\frac{1}{\beta}\right)
{\hbox{e}}^{-\gamma(t-t')}-\frac{1}{\gamma-\beta/m}{\hbox{e}}^{-\beta(t-t')/m}. 
\end{equation}

The characteristic function of $p(x,t)$ directly follows from Eq.(\ref{f2}) \cite{doo,west}: 
\begin{equation}
\label{ave}
{\widetilde p}(k,t)=\langle{\hbox{e}}^{ikx(t)}\rangle=
{\hbox{e}}^{ikf_1}\langle\exp(ik\int_0^tf_2(t')dt')\rangle=
{\hbox{e}}^{ikf_1}\exp\left(-K^\alpha|k|^\alpha\int_0^t|f_2(t')|^\alpha dt'\right). 
\end{equation}
Eq.(\ref{ave}) implies the L\'evy distribution with the same stability 
index as the driving noise $L$ and a translation parameter which coincides with $f_1$ 
and for large time equals either $f_0/\lambda$ ($\lambda\ne0$) or 
$f_0(t-m/\beta)$ ($\lambda=0$). Since the dependence of the density 
distribution on $f_0$ is trivial, we assume in the following $f_0=0$ 
and $v_0=0$. Then the distribution is symmetric for any time. 
The inverse Fourier transform can be conveniently expressed in a form 
of the Fox function \cite{mat,sri}: 
\begin{eqnarray}
\label{pfok}
p(x,t)=Nf(t) H_{2,2}^{1,1}\left[f(t) |x|\left|\begin{array}{c}
(1-1/\alpha,1/\alpha),(1/2,1/2)\\
\\
(0,1),(1/2,1/2)
\end{array}\right.\right], 
  \end{eqnarray} 
where $f(t)=K^\alpha\int_0^t|f_2(t')|^\alpha dt'$. By introducing 
a new variable $\tau=t-t'$ we have $f(t)=K^\alpha(\int_0^T+\int_T^t)$ 
and the exponentials in the second integral can be dropped for any $T\gg1$. 
Therefore $f(t)=t/\beta+$const for large times. Since $p(x,t)\sim |x|^{-1-\alpha}$ 
for $|x|\to\infty$, variance and all higher moments 
of the distribution (\ref{pfok}) are divergent, 
as well as the average if $\alpha\le1$. A relative expansion rate can 
be quantified by the fractional moments of the order $\delta<\alpha$, 
$\langle|x|^\delta\rangle(t)$; they are given by the Mellin transform from 
the Fox function \cite{kla}. The final expression reads 
\begin{equation}
\label{mom}
\langle|x|^\delta\rangle=\frac{2}{\alpha}f(t)^{\delta/\alpha}
\frac{\Gamma(-\delta/\alpha)\Gamma(1+\delta)}{\Gamma(-\delta/2)\Gamma(1+\delta/2)}. 
\end{equation}
Consequently, in the limit of large $t$ the fractional moments decrease 
with the damping coefficient $\beta$ and the expansion rate is large for 
small $\alpha$. 

As an example, let us consider the case $\alpha=1$ for which 
results take the transparent form. 
This particular value of the stability parameter corresponds to the well-known 
Cauchy distribution; it was considered in Ref.\cite{garb} for the white noise. 
For $\gamma>\beta/m$, a straightforward calculation yields the expression 
for the apparent width of the distribution $p(x,t)$: 
\begin{equation}
\label{fcau}
f(t)=\frac{t}{\beta}-\frac{1+m}{\gamma\beta}-\frac{1}{\gamma}
\left(\frac{1}{\gamma-\beta/m}-\frac{1}{\beta}\right)
{\hbox{e}}^{-\gamma t}+\frac{m}{\beta}\frac{1}{\gamma-\beta/m}{\hbox{e}}^{-\beta t/m}. 
\end{equation}
In the limit of the large time, inertia and noise relaxation time are responsible 
for a time-shift which is negative and rises with $m$ and $1/\gamma$.

\section{Multiplicative noise} 

By introducing a multiplicative noise we take into account that the influence 
of the random component of the dynamics depends on the dynamics itself. 
The one-dimensional case is given by the Langevin equation 
\begin{eqnarray}
\label{lam}
m\dot v(t)&=&-\partial V(x)/\partial x-\beta v(t)+\gamma G(x)\xi(t)\nonumber\\
\dot x(t)&=&v(t)\nonumber\\
d\xi(t)&=&-\gamma\xi(t)dt+dL(t). 
\end{eqnarray}
In the following we assume the noise intensity in the algebraic form, 
$G(x)=|x|^{-\theta/\alpha}$ $(\alpha+\theta>0)$. 

The simplest problem involves the white noise ($\gamma\to\infty$) and 
neglects the inertia. The latter condition means that mass is small compared to 
the damping parameter $\beta$ and the strength of the noise. The case 
of the normal distribution, $\alpha=2$, is well-known \cite{schen,vkam,gra}. 
A stochastic integral in the Langevin equation is not completely determined for the 
uncorrelated noise, since it is not clear whether the dynamical variable in 
the function $G(x)$ should be evaluated at the time before the noise acts, 
after that, or somewhere in between. Two interpretations are of particular 
importance. In the It\^o interpretation, the $i$'s component of the discretized 
stochastic integral is $G[x(t_{i-1})][\xi(t_i)-\xi(t_{i-1})]$, whereas 
the Stratonovich interpretation includes both the beginning and the end 
of each interval: $G\{[x(t_{i-1})+x(t_i)]/2\}[\xi(t_i)-\xi(t_{i-1})]$. Both 
assumptions result in a different Fokker-Planck equation but a difference 
resolves itself solely to a drift term (the spurious drift) which can be 
eliminated by an appropriate modification of the deterministic potential. 
For this reason, a physical relevance of the It\^o-Stratonovich dilemma 
is disputed \cite{vkam}. Nevertheless, physical implications of both 
interpretations may be different. For example, phase transitions due to 
instabilities of the disordered phase in the framework of the 
Ginzburg-Landau model take place only for the Stratonovich 
interpretation \cite{car}. Rules of the ordinary calculus are valid in 
the Stratonovich formalism, in contrast to the It\^o interpretation. 

The meaning of both interpretations becomes more transparent when we 
take into account the finite correlations and inertia. First of all, 
the memory effects favour the Stratonovich interpretation since it constitutes 
the white-noise limit of the correlated processes \cite{won}. Inertia acts 
in the opposite direction. It was demonstrated \cite{kup} -- by the estimation 
of the velocity moments and using the It\^o formula -- that if the inertia 
relaxation time goes to zero faster than the noise correlation time, the
Stratonovich interpretation is valid. The opposite limit produces the It\^o result. 
If both time scales are comparable, neither of the above interpretations is valid. 
We will demonstrate that similar conclusions can be drawn for the general L\'evy 
stable processes. However, for $\alpha<2$ methods of Ref.\cite{kup} cannot 
be applied since the moments are divergent and the It\^o formula is unknown. 

We take into account only It\^o and Stratonovich interpretations of the stochastic integral. 
The It\^o interpretation applies, beside the systems with large mass, to discrete problems; 
it is commonly used in the perturbation theory \cite{gar}. However, there are 
indications that other interpretations are also important. 
For example, it was recently experimentally demonstrated that description of the Brownian 
motion in the presence of gravitational and electrostatic forces requires a backward 
integral (anti-It\^o interpretation) \cite{volpe}. 

\subsection{White-noise case without the inertia} 

To study a diffusion process, we consider a free particle, $V(x)=0$. In the limit 
$m\to 0$ and $\gamma\to\infty$, Eq.(\ref{lam}) becomes a single Langevin equation 
of the first order: 
\begin{equation}
\label{lam0}
dx(t)=|x(t)|^{-\theta/\alpha}dL(t),
\end{equation}
where, for simplicity, we assumed $\beta=1$. In the It\^o interpretation it corresponds 
to the Fokker-Planck equation \cite{sche} 
\begin{equation}
\label{fraceo}
\frac{\partial}{\partial t}p_I(x,t)=
K^\alpha\frac{\partial^\alpha}{\partial |x|^\alpha}[|x|^{-\theta}p_I(x,t)], 
\end{equation}
which differs from the equation for the additive noise 
by an algebraic term under the fractional derivative. This particular form of the 
multiplicative factor suggests simple scaling properties and a possible similarity 
of the solution to Eq.(\ref{pfok}). Indeed, an asymptotic solution of Eq.(\ref{fraceo}) 
can be found in the form $a(t)H_{2,2}^{1,1}[a(t)x]$. The 
procedure is the following \cite{physa}. First, we insert the above expression 
to the Langevin equation and take the Fourier transform, which also has a form of 
the Fox function but of a higher order. Expansion of the Fox functions in powers 
of $k$ and neglecting the terms of the order $|k|^{2\alpha+\theta}$ and higher 
yields a simple differential equation for the function $a(t)$ and allows us to determine 
some Fox function coefficients. Finally, we obtain the solution 
\begin{eqnarray}
\label{pfokm}
p_I(x,t)=Na(t) H_{2,2}^{1,1}\left[a(t) |x|\left|\begin{array}{c}
(1-\frac{1-\theta}{\alpha+\theta},\frac{1}{\alpha+\theta}),(a_2,A_2)\\
\\
(b_1,B_1),(1-\frac{1-\theta}{2+\theta},\frac{1}{2+\theta})
\end{array}\right.\right], 
  \end{eqnarray} 
where $a(t)\sim t^{-1/(\alpha+\theta)}$ and the coefficients $(a_2,A_2)$ and 
$(b_1,B_1)$ are arbitrary. The asymptotic form of the solution is the same 
as for the driving noise, 
\begin{equation}
\label{pi}
p_I(x,t)\sim|x|^{-1-\alpha}. 
\end{equation}
Therefore, for the It\^o interpretation the variance is always divergent 
which implies accelerated diffusion. Fractional moments can be evaluated 
similarly to the additive noise case; a straightforward calculation yields 
\begin{equation}
\label{mommi}
\langle|x|^\delta\rangle\sim t^{\delta/(\alpha+\theta)}.
\end{equation}
The multiplicative noise parameter $\theta$ 
can both strengthen ($\theta<0$) and weaken ($\theta>0$) 
the time-dependence of $\langle|x|^\delta\rangle$, compared to the case 
of the additive noise. The undetermined coefficients do not influence 
the functional dependences. The same method of solution can be applied in 
the presence of the linear deterministic force $F(x)=-\lambda x$ \cite{sro1}. 

Results for the Stratonovich interpretation are qualitatively different 
since the decline of the noise intensity with $x$ may compensate the effect 
of the long jumps. The technical advantage of this interpretation, for one-dimensional 
systems, consists in a possibility of applying rules of the ordinary calculus. 
This property is strict for $\alpha=2$ \cite{gar}. In the general case, 
the noise distribution must be truncated, a requirement that is obvious 
for the linear systems \cite{sro2}. However, it was numerically 
demonstrated that in practice cases with the distribution without 
any cut-off also comply with rules of the ordinary calculus if the system is 
nonlinear \cite{sro1,sro2}. Then we may define a new variable, 
\begin{equation}
\label{yodx}
y(x)=\frac{\alpha}{K(\alpha+\theta)}|x|^{1+\theta/\alpha}\hbox{sgn}(x), 
\end{equation}
which transforms Eq.(\ref{lam0}) to the equation with the additive noise. 
The asymptotic form of the solution \cite{sro1}
\begin{equation} 
\label{sas}
p_S(x,t)\sim t^{\alpha/(\alpha+\theta)}|x|^{-1-\alpha-\theta}~~~~(|x|\to\infty)
\end{equation}
implies that variance may be convergent. It takes the form 
$\langle x^2\rangle\sim t^{2/(\alpha+\theta)}$
on the condition $\alpha+\theta>2$. Therefore, diffusion is either 
anomalously weak -- if the above condition is satisfied -- or accelerated \cite{uwa}. 

\subsection{General case} 

\begin{center}
\begin{figure}
\includegraphics[width=12cm]{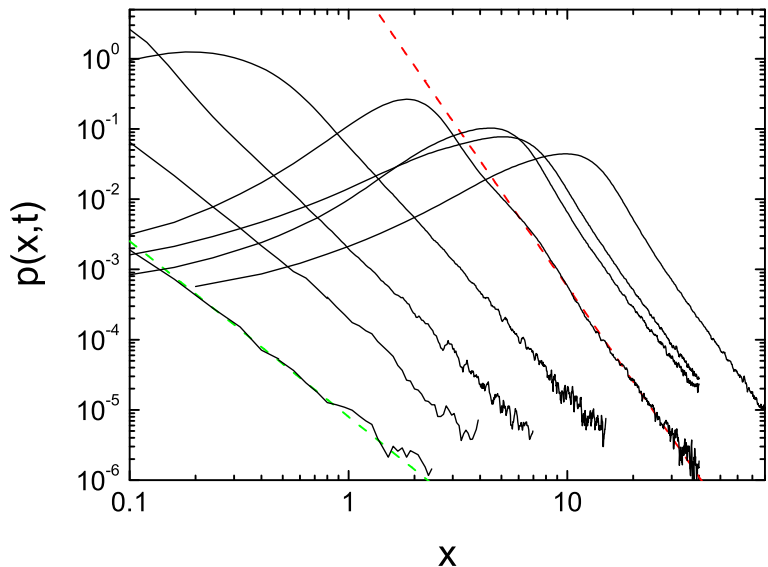}
\caption{(Colour online) Probability density distributions at $t=1$ for $\alpha=1.5$, $\theta=2$, 
$\gamma=100$ and $\beta=1$. The curves correspond to the following values of $m$ 
(from left to right): $5\times10^5$, $5\times10^4$, $10^4$, $10^3$, 0.01, 0.1, 10, 1.
The red dashed line marks the depedence $x^{-4.5}$ and the green dashed line (at the left side) 
the dependence $x^{-2.5}$. Each curve was obtained by averaging over $10^8$ trajectories.}
\end{figure}
\end{center}

First let us consider the overdamped limit (the adiabatic approximation) 
by putting $m=0$ in Eq.(\ref{lam}). Equations take the form 
\begin{eqnarray}
\label{labezm}
\dot x(t)&=&\frac{\gamma}{\beta} |x|^{-\theta/\alpha}\xi(t)\nonumber\\
d\xi(t)&=&-\gamma\xi(t)dt+dL(t). 
\end{eqnarray}
After transformation of the process variable according to Eq.(\ref{yodx}), 
we obtain a linear equation with the additive noise. Its solution reads 
$y(t)=\int_0^t f_2(t') L(t')dt'$, where $f_2(t')=(1-{\hbox{e}}^{-\gamma(t-t')})/\beta$, 
and the density distribution of $y$ has the L\'evy form, Eq.(\ref{pfok}). 
Transformation to the variable $x$ yields the final result: 
\begin{eqnarray} 
\label{solsx0}
p(x,t)=\frac{\alpha+\theta}{\alpha^2|x|}H_{2,2}^{1,1}\left[\frac{|x|^{1+\theta/\alpha}}
{K(1+\theta/\alpha)f(t)^{1/\alpha}}
\left|\begin{array}{l}
(1,1/\alpha),(1,1/2)\\
\\
(1,1),(1,1/2)
\end{array}\right.\right], 
\end{eqnarray}
where $f(t)=K^\alpha\int_0^t|f_2(t')|^\alpha dt'$. The expansion of Eq.(\ref{solsx0}) 
in the fractional powers of $|x|^{-1}$ yields an approximation of the solution 
for large $|x|$ and the first term is of the form 
\begin{equation} 
\label{sasc}
p(x,t)\sim f(t)^{\alpha/(\alpha+\theta)}|x|^{-1-\alpha-\theta}. 
\end{equation}
If $\alpha+\theta>2$, the variance is convergent and it can be exactly evaluated 
by using properties of the Fox functions, in particular, an expression for 
the Mellin transform. A straightforward calculation yields  
\begin{equation}
\label{var}
\langle x^2\rangle=
-\frac{2}{\pi\alpha}\left[K(\frac{\theta}{\alpha}+1)\right]^{2\alpha/(\alpha+\theta)}
\Gamma\left(-\frac{2}{\alpha+\theta}\right)
\Gamma\left(1+\frac{2\alpha}{\alpha+\theta}\right)\sin\left(\frac{\pi\alpha}{\alpha+\theta}\right)
f(t)^{2/(\alpha+\theta)}. 
\end{equation}
Eq.(\ref{solsx0}) converges to the white-noise solution in the Stratonovich 
interpretation for any noise relaxation parameter $\gamma$ if time is large or, 
for any time, if $\gamma\to\infty$. 
\begin{center}
\begin{figure}
\includegraphics[width=12cm]{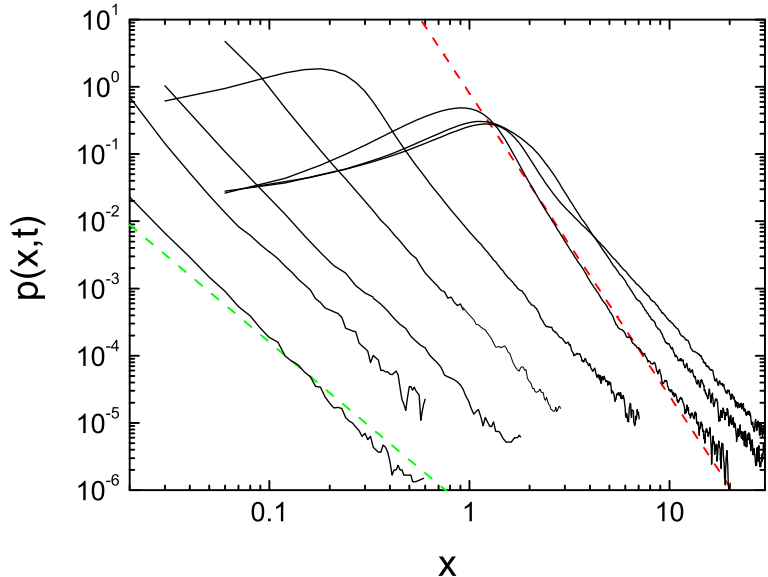}
\caption{(Colour online) Same as Fig.1 but for $\gamma=1$. The following values of $m$ 
are presented (from left to right): $5\times10^5$, $5\times10^4$, $10^4$, $10^3$, 
$10^2$, 0.01, 0.1, 1.}
\end{figure}
\end{center}

\begin{center}
\begin{figure}
\includegraphics[width=12cm]{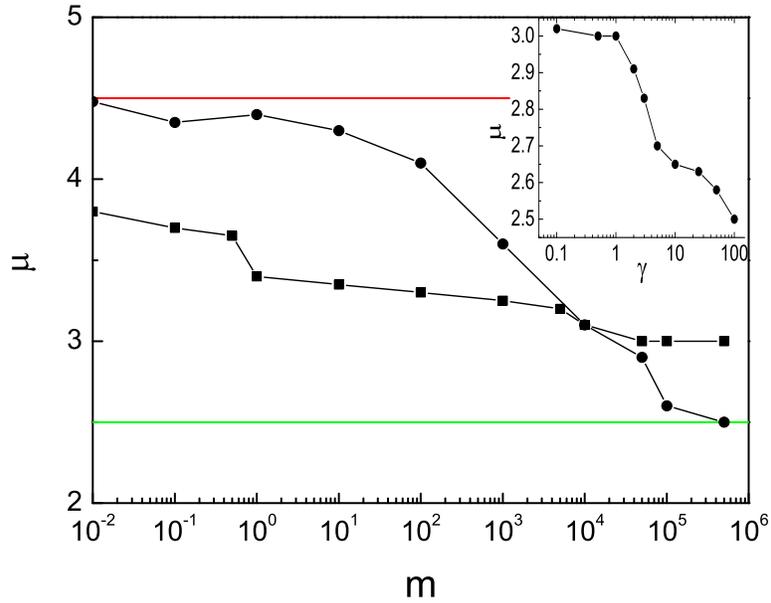}
\caption{(Colour online) Slopes of the tails, $|x|^{-\mu}$, of $p(x,t)$ at 
$t=1$ for $\alpha=1.5$, $\theta=2$ and $\beta=1$, as a function of $m$. Two cases are presented: 
$\gamma=100$ (points) and $\gamma=1$ (squares). The limiting values, $\mu_S$ and $\mu_I$, 
are marked by the horizontal lines. Inset: $\mu$ as a function of $\gamma$ for $m=5\times 10^5$. 
}
\end{figure}
\end{center}

Distributions for arbitrary $\gamma$ and $m$ have been obtained by a numerical 
integration of the stochastic equations, Eq.(\ref{lam}). For that purpose, a second 
order difference approximation, called a St\"ormer method \cite{dahl}, was applied to the 
first two equations. Since the resulting difference equations are implicit, the parabolic 
interpolation scheme was applied at each step \cite{ral}. The third equation was integrated 
by an Euler method and the noise term in the $i$-step was represented by $\tau^{1/\alpha}L_i$, 
where $\tau$ was a time step \cite{wer}. We will demonstrate how the asymptotic shape 
of the distribution, for a given time, depends on $m$ and $\gamma$. Another quantity 
of interest is a time dependence of the variance. 

Fig.1 presents the probability density distributions as a function of the particle mass 
in the limit of the white noise at $t=1$. The distributions widen with $m$ up to $m=1$ but 
then the trend goes into reverse. For the large mass the distributions have a form 
of the delta function accompanied by a little tail. The tails are algebraic, 
$\sim|x|^{-\mu}$, and the slope $\mu$ diminishes with the mass. Two limiting 
values, $\mu_S=\alpha+\theta+1$ and $\mu_I=\alpha+1$, correspond to the Stratonovich, 
Eq.(\ref{sas}), and It\^o, Eq.(\ref{pi}), interpretations, respectively. 
The distributions for the case of the finite noise relaxation time 
($\gamma=1$) are presented in Fig.2. They are similar to those for the white 
noise but the limiting slopes are not yet reached at $t=1$. Slopes for all cases 
are put together in Fig.3. The dependence 
$\mu(m)$ is flat for $\gamma=1$ whereas for the white-noise case large values of $\mu$ 
dominate and there is a rapid transition to $\mu_I$. Variance is finite ($\mu>3$) 
except for the very large $m$. This case is separately presented in Fig.3: $\mu$ rises 
with the noise relaxation time from the It\^o value for the white noise, $\mu=2.5$, 
to $\mu=3$, where it saturates. 

\begin{center}
\begin{figure}
\includegraphics[width=12cm]{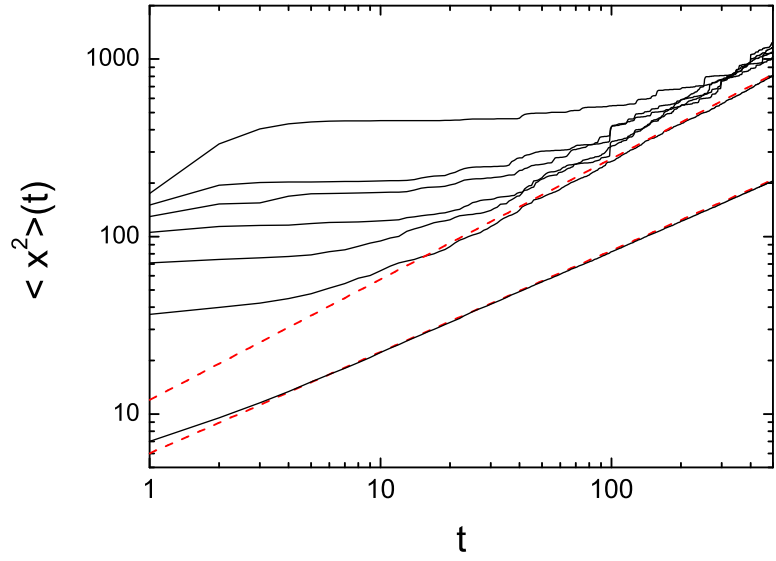}
\caption{(Colour online) Variance as a function of time for $\alpha=1.5$, $\theta=2$, 
$\gamma=100$ and $\beta=1$. The curves correspond to the following values of $m$ 
(from bottom to top): 0.01, 0.1, 0.2, 0.3, 0.4, 0.5, 1.
The red dashed lines mark the dependence $t^{0.57}$ (lower) and $t^{0.68}$ (upper).}
\end{figure}
\end{center}

\begin{center}
\begin{figure}
\includegraphics[width=12cm]{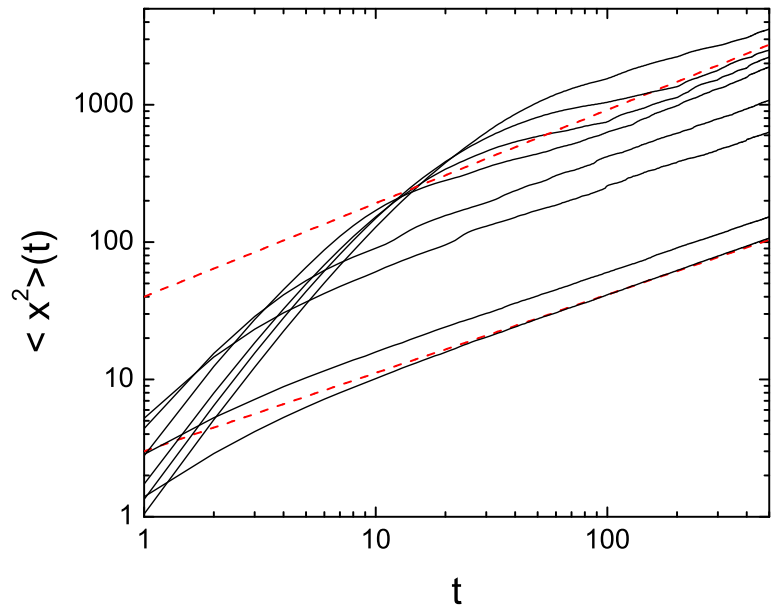}
\caption{(Colour online) The same as Fig.4 but for $\gamma=1$. 
The curves correspond to the following values of $m$ 
(from bottom to top on the right side): 0.01, 0.1, 1, 2, 5, 10, 15, 20.}
\end{figure}
\end{center}

Diffusion properties of the system are determined by a long-time behaviour of 
the variance. The case corresponding to the short noise relaxation time 
is presented in Fig.4. If $m$ is very small, the variance assumes the form 
$\langle x^2\rangle\sim t^{2/(\alpha+\theta)}$ for the large time. The slope becomes 
slightly larger if $m$ is not infinitesimal; 
it equals 0.68 for all $m\ge0.1$. Therefore, all cases 
indicate a sub-linear time dependence for the large time: the diffusion process 
is anomalously weak (subdiffusion). On the other hand, if time is not very large, 
$\langle x^2\rangle(t)$ exhibits a plateau which widens with $m$. Moreover, the curves 
reveal a stepwise pattern which can be attributed to 
a competition between the expansion and the attraction to the origin. Such a behaviour 
of the curves in Fig.4 is a clear consequence of the lack of memory. For $\gamma=1$ 
the dependence $\langle x^2\rangle(t)$ -- presented in Fig.5 -- is smooth; it assumes 
the asymptotic shape $t^{0.68}$ for large $m$, similar to the previous case. 
If $m$ is close to zero, variance is given by Eq.(\ref{var}). 

Also properties of more complicated systems are modified when we take into account 
the finite relaxation time and inertia. Let us consider the following potential 
\begin{equation}
\label{potb}
V(x)=\frac{A}{4}x^4-\frac{B}{2}x^2 
\end{equation}
which has the double well shape. The mean first passage time (MFPT) 
is a quantity of particular importance \cite{han}; it was studied in the context of 
the L\'evy stable processes in Ref.\cite{dit,dyb2}. The case of the multiplicative noise 
was discussed in Ref.\cite{sro2}; it was demonstrated that MFPT decreases with 
$\theta$ but the rate depends on the particular interpretation of the stochastic 
integral. In this paper we calculate MFPT for finite $m$ and $\gamma$ by integration 
of Eq.(\ref{lam}) with the absorbing barriers at $x(0)=-\sqrt{B/A}$ 
and $x=0$. The latter boundary condition is nonlocal due to the jumps \cite{dyb2}. 
The resulting MFPT, as a function of $\theta$, is presented in Fig.6. All the curves 
fall since the effective depth of the potential decreases with $\theta$. MFPT rises with 
$m$, because of the increasing attraction to the origin, 
and becomes flat. A similar effect 
is observed for the decreasing $\gamma$ (stronger memory) since then the intensity of 
the driving noise is smaller. In the white noise limit, $\gamma\to\infty$, 
the Stratonovich result is recovered. 
\begin{center}
\begin{figure}
\includegraphics[width=12cm]{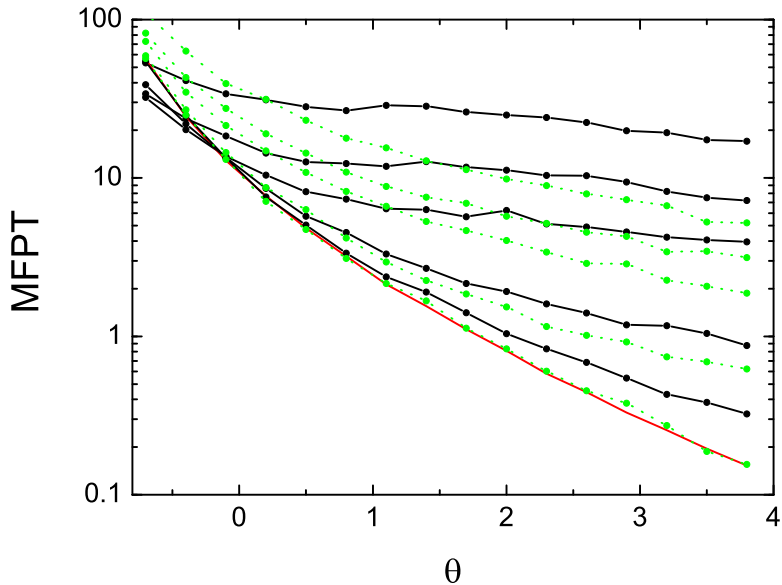}
\caption{(Colour online) MFPT as a function of $\theta$ for finite $m$ and $\gamma$. 
The black solid lines correspond to the following values of $m$ 
(from bottom to top): 0.001, 0.01, 0.1, 1, 10 and $\gamma=100$. Dependence on 
$\gamma$ for $m=0$ is marked by green dotted lines: $\gamma=1000$, 10, 1, 0.5, 0.2 
(from bottom to top). Other parameters: $\alpha=1.5$, $\beta=1$, $A=1$ and $B=0.1$. 
Result for the Stratonovich interpretation is marked by the red solid line without symbols.}
\end{figure}
\end{center}

\section{Summary and conclusions}

We have studied a one-dimensional dynamics of a massive particle subjected to 
the general L\'evy stable noise, both additive and multiplicative. 
The driving noise has been represented by the generalised Ornstein-Uhlenbeck 
process and then the finite noise relaxation time has been taken into account. 
In the linear case, the dynamical variable $x$ is governed by the L\'evy 
distribution and the parameter $\alpha$ is the same 
as for the driving noise. Therefore, diffusion is always accelerated. 
Distribution converges with time to the white-noise and 
massless case. Fractional moments rise with time; the rate decreases with 
the stability index $\alpha$ and the damping coefficient $\beta$. Inertia and 
noise relaxation time influence the rate of convergence to the asymptotic 
distribution. 

The $x$-dependence of the multiplicative noise modifies the distribution. 
Slopes of the tail depend on the multiplicative factor $G(x)$, which was 
assumed in the algebraic form, and variance is finite if $G(x)$ 
falls sufficiently fast. Variance rises sub-linearly with time for $t\gg1$ 
which indicates the subdiffusion. Those conclusions are valid for any noise 
relaxation parameter $\gamma$. The limit $\gamma\to\infty$ is of particular 
importance; the distribution in this limit coincides 
with Eq.(\ref{sas}). Therefore, the limit of the Langevin equation driven by 
the generalised Ornstein-Uhlenbeck process produces the same result as 
the formal variable change in the Langevin equation for the white-noise case. 
The influence of inertia is more subtle. 
It favours an expansion of the distribution if $m$ is small but for large $m$ 
distribution shrinks to the delta function. However, even in the limit $m\to\infty$ 
a little tail remains and it makes the variance divergent. In the white-noise 
limit, that tail agrees with the distribution in the It\^o interpretation. 
On the other hand, the Stratonovich interpretation is valid for the small mass. 
Those conclusions are similar to the case of the normal distribution \cite{kup}. 
Since slowly falling tails have been encountered only for the extremely large 
masses, convergent variance is by no means exceptional for the L\'evy stable 
processes: it emerges if intensity of the multiplicative noise diminishes 
sufficiently fast. The finite noise relaxation time and inertia affect the barrier 
penetration: the calculated MFPT rises with both the memory parameter $1/\gamma$ and 
the particle mass. In the white-noise limit MFPT converges to the Stratonovich result. 

The above analysis demonstrates that the Langevin formalism with the 
multiplicative L\'evy noise predicts heavy, algebraic tails of the probability 
density distribution and the index $\mu$ can assume arbitrarily large values. 
As a consequence, moments of an arbitrarily high order may be convergent. 
$\mu$ depends not only on $\alpha$ and $\theta$, as it is the case for the massless 
particle, but also on the inertia. Those conclusions suggest that the presented formalism 
may be well suited to describe processes characterised by a variety of the algebraic slopes 
of the distribution \cite{sta,gab,sche}. In the field of finance, 
a traditional Black-Scholes model of option pricing, which includes the additive Gaussian 
noise, can be generalised by introducing the L\'evy flights. Need of such a generalisation 
is obvious \cite{cart} but, since variance of the additive L\'evy process is infinite, 
a truncation of the distribution becomes necessary. On the other hand, the first-order 
equation, like the Black-Scholes equation, with the multiplicative noise 
predicts sufficiently steep distribution slopes to ensure 
the finite variance also without any truncation if the stochastic integral is understood in 
the Stratonovich sense. The present paper justifies this interpretation for the first-order 
stochastic equations: it demonstrates that distribution slopes are robust 
in respect to the noise relaxation time -- which is always finite for realistic problems -- 
and the white-noise limit exists.


\begin{thebibliography}{99}

\bibitem{sta}
H. E. Stanley, Physica A {\bf 318}, 279 (2003). 

\bibitem{gab}
X. Gabaix, P. Gopikrishnan, V. Plerou, and H. E. Stanley, Nature {\bf 423}, 267 (2003). 

\bibitem{sche}
D. Schertzer, M. Larchev\^{e}que, J. Duan, V. V. Yanovsky, and S. Lovejoy, 
J. Math. Phys. {\bf 42}, 200 (2001). 

\bibitem{che}
A. Chechkin, V. Gonchar, J. Klafter, R. Metzler, and L. Tanatarov, 
Chem. Phys. {\bf 284}, 233 (2002). 

\bibitem{kla}
R. Metzler and J. Klafter, Phys. Rep. {\bf 339}, 1 (2000).

\bibitem{srkan}
T. Srokowski, Physica A {\bf 390}, 3077 (2011).

\bibitem{newm}
M. E. J. Newman, Contemp. Phys. {\bf 46}, 323 (2005). 

\bibitem{park}
M. Park, N. Kleinfelter, and J. H. Cushman, 
Phys. Rev. E {\bf 72}, 056305 (2005).

\bibitem{osh}
B. O'Shaughnessy and I. Procaccia, Phys. Rev. Lett. {\bf 54}, 455 (1985).

\bibitem{met4}
R. Metzler and T. F. Nonnenmacher, J. Phys. A {\bf 30}, 1089 (1997).

\bibitem{bro}
D. Brockmann and T. Geisel, Phys. Rev. Lett. {\bf 90}, 170601 (2003).

\bibitem{manor}
A. Manor and N. M. Shnerb, Phys. Rev. Lett. {\bf 103}, 030601 (2009). 

\bibitem{cogn}
A. La Cognata, D. Valenti, A. A. Dubkov, and B. Spagnolo, 
Phys. Rev. E {\bf 82}, 011121 (2010). 

\bibitem{sro1}
T. Srokowski, Phys. Rev. E {\bf 80}, 051113 (2009). 

\bibitem{hae}
P. H\"anggi and P. Jung, 
in {\it Advances in Chemical Physics}, edited by I. Prigogine and S. A. Rice, 
vol. 89 (John Wiley \& Sons, Inc., Hoboken, NJ, USA, 2007).

\bibitem{ter}
J. N. Teramae, H. Nakao, and G. B. Ermentrout, Phys. Rev. Lett. {\bf 102}, 194102 (2009). 

\bibitem{sho}
R. Short, L. Mandel, and R. Roy, Phys. Rev. Lett. {\bf 49}, 647 (1982). 

\bibitem{kub}
R. Kubo, {\it Fluctuation, Relaxation and Resonance in Magnetic Systems} 
(Oliver and Boyd, London, 1982). 

\bibitem{mang}
S. E. Mangioni, R. R. Deza, R. Toral, and H. S. Wio, Phys. Rev. E {\bf 61}, 223 (2000).

\bibitem{won}
E. Wong and M. Zakai, Ann. Math. Stat. {\bf 36}, 1560 (1965). 

\bibitem{kup}
R. Kupferman, G. A. Pavliotis, and A. M. Stuart, Phys. Rev. E {\bf 70}, 036120 (2004).

\bibitem{san}
J. M. Sancho and A. Sanchez, Eur. Phys. J. B {\bf 16}, 127 (2000).

\bibitem{west}
B. J. West and V. Seshardi, Physica A {\bf 113}, 203 (1982).

\bibitem{jes}
S. Jespersen, R. Metzler, and H. C. Fogedby, 
{\it Phys. Rev.} {\bf E59}, 2736 (1999). 

\bibitem{garb}
P. Garbaczewski and R. Olkiewicz, J. Math. Phys. {\bf 41}, 6843 (2000). 

\bibitem{yan}
V. V. Yanovsky, A. V. Chechkin, D. Schertzer, and A. V. Tur,
Physica A {\bf 282}, 13 (2000). 

\bibitem{barkai}
E. Barkai, {\it Phys. Rev.} {\bf E68}, 055104 (2003). 

\bibitem{emb}
P. Embrechts and M. Maejima, {\it Selfsimilar Processes} (Princeton University Press, Princeton, 2002). 

\bibitem{elia}
I. Eliazar and J. Klafter, Physica A {\bf 376}, 1 (2007). 

\bibitem{act}
T. Srokowski, Acta Phys. Polon. B {\bf 42}, 3 (2011). 

\bibitem{man}
R. N. Mantegna and H. E. Stanley, Phys. Rev. Lett. {\bf 73}, 2946 (1994).

\bibitem{doo}
J. L. Doob, Ann. Math. {\bf 43}, 351 (1942). 

\bibitem{mat}
A. M. Mathai and R. K. Saxena, {\it The $H$-function with Applications in Statistics 
and Other Disciplines} (Wiley Eastern Ltd., New Delhi, 1978).

\bibitem{sri}
H. M. Srivastava, K. C. Gupta, and S. P. Goyal, 
{\it The $H$-functions of one and two variables with applications} 
(South Asian Publishers, New Delhi, 1982). 

\bibitem{schen}
A. Schenzle and H. Brand, Phys. Rev. A {\bf 20}, 1628 (1979). 

\bibitem{vkam}
N. G. van Kampen, J. Stat. Phys. {\bf 24}, 175 (1981). 

\bibitem{gra}
R. Graham and A. Schenzle, Phys. Rev. A {\bf 25}, 1731 (1982). 

\bibitem{car}
O. Carrillo, M. Iba\~nes, J. Garc\'ia-Ojalvo, J. Casademunt, and J. M. Sancho, 
Phys. Rev. E {\bf 67}, 046110 (2003).

\bibitem{gar}
C. W. Gardiner, {\it Handbook of Stochastic Methods for Physics, Chemistry
and the Natural Sciences} (Springer-Verlag, Berlin, 1985).

\bibitem{volpe}
G. Volpe, L. Helden, T. Brettschneider, J. Wehr, and C. Bechinger, 
Phys. Rev. Lett. {\bf 104}, 170602 (2010). 

\bibitem{physa}
T. Srokowski, Physica A {\bf 388}, 1057 (2009). 

\bibitem{sro2}
T. Srokowski, Phys. Rev. E {\bf 81}, 051110 (2010). 

\bibitem{uwa}
In contrast to the case $\alpha<2$, the shape of the density distribution for 
$\alpha=2$ does not depend on the specific interpretation of the stochastic integral. 
For the It\^o case, the exact solution of the corresponding Fokker-Planck equation 
predicts a stretched-Gaussian shape, $p_I(x,t)\sim \exp(-{\hbox{const}}|x|^{2+\theta}/t)$, 
for large $x$ (H. G. E. Hentschel and I. Procaccia, Phys. Rev. A {\bf 29}, 1461 (1984)). 
The same form follows when we apply the transformation (\ref{yodx}) to the Gaussian. 

\bibitem{dahl}
A. Bj\"ork and G. Dahlquist, {\it Numeriska Metoder} (Liber Grafiska AB, Stockholm, 1969). 

\bibitem{ral}
A. Ralston, {\it A First Course in Numerical Analysis} (McGraw-Hill, New York, 1965). 

\bibitem{wer}
A. Janicki and A. Weron, {\it Simulation and Chaotic Behavior of
$\alpha$-Stable Stochastic Processes} (Marcel Dekker, New York, 1994). 

\bibitem{han}
P. H\"anggi, P. Talkner, and M. Borkovec, Rev. Mod. Phys. {\bf 62},
251 (1990). 

\bibitem{dit}
P. D. Ditlevsen, Phys. Rev. E {\bf 60}, 172 (1999). 

\bibitem{dyb2}
B. Dybiec, E. Gudowska-Nowak, and P. H\"anggi, 
Phys. Rev. E {\bf 75}, 021109 (2007). 

\bibitem{cart}
A. Cartea and D. del-Castillo-Negrete, Physica A {\bf 374}, 749 (2007). 

\end{thebibliography}
\end{document}